# Unravelling the role of waveguide-plasmon strong coupling, avoided crossing and natural weak value amplification on the enhanced Faraday effect in hybrid magneto-plasmonic systems

Jeeban K Nayak[+], Shyamal Guchhait[+], Ankit K Singh and Nirmalya Ghosh*

*Department of Physical Sciences,*

*Indian Institute of Science Education and Research (IISER) Kolkata.*

*Mohanpur 741246, India*

+these authors contributed equally

**Corresponding author:nghosh@iiserkol.ac.in*

**Abstract**

Enhancement of magneto-optical effects in hybrid magneto-plasmonic systems has attracted considerable recent attention owing to their potential for building non-reciprocal nanophotonic devices. Quantitative understanding on the fundamental origin and contributing mechanisms for the enhancement is extremely crucial for optimizing applications. Here, we unravel the distinctly different physical origin of the giant enhancement of Faraday rotation and ellipticity in a hybrid magneto-plasmonic system, namely, waveguided magneto-plasmonic crystal for excitation with transverse electric (TE) and transverse magnetic (TM) polarized light. With TE polarization excitation, where the surface plasmons are not directly excited, the natural weak value amplification of Faraday effects appearing due to the spectral domain interference of Fano resonance is the dominant cause of the enhancement. For TM polarization excitation, on the other hand, waveguide-plasmon strong coupling and its universal manifestation of avoided crossing plays an intriguing role, leading to maximum enhancement of the magneto-optical effects in the avoided crossing regime.

## Introduction

Hybridization of magneto-optic (MO) materials with plasmonic nanostructures enables magnetic-tuneable multifunctional nano-devices that provides precise control over its plasmonic and magnetic properties[1,2]. Moreover, such systems are also known to exhibit giant enhancement of the MO effects (Faraday, Kerr effect etc.)[2–8]. Note that the non-reciprocal nature of the MO effects is extremely crucial for developing optical isolators[3], modulators, and optical-magnetic data storage devices[1,8] etc. The very weak MO effects of the available materials are a major stumbling block towards their applications in nano scale devices. Significant efforts have therefore been delivered in the last few years to demonstrate enhancement of both the magneto-optical Kerr effect and the Faraday effect in magneto-plasmonic nanostructures[2,4,7]. Earlier approaches for increasing the MO activity of the systems were mostly based on the excitation of surface plasmons, where the near field enhancement associated with the surface plasmons leads to substantial enhancement of MO responses of samples[9,10]. In this regard, a recent study has attracted a lot of attention that has demonstrated giant enhancement of Faraday rotation in a waveguided magneto-plasmonic crystal (WMPC)[3] system. Some other related studies have reported even larger enhancement of Faraday rotation in various magneto-plasmonic systems using similar approach[4,6,11]. In most of these systems, the enhancement of MO effects is attributed to either plasmon-induced near field enhancement or the coupling of waveguide-plasmon-polariton[2,4,6]. In either of the scenarios, the MO effect enhancement is crucially related to the excitation of the surface plasmon resonance, which in case of the investigated planar structures can only be excited using transverse magnetic (TM) polarization of light. In contrast, in our recent investigation a large enhancement of the Faraday rotation and ellipticity was shown in a Fano resonant WMPC for input TE polarization where no surface plasmons were directly excited[12]. This observation initially indicated towards another important mechanism of the enhancement, namely, the natural weak value amplification (WVA) of Faraday rotation and ellipticity due to spectral domain interference of Fano resonance. Proper understanding of the physical origin of the enhanced MO effects in hybrid magneto-plasmonic system and its dependence on the various physically accessible parameters of the system and excitation light is of paramount importance for both fundamental and applied interests.

The concept of weak measurement and WVA was introduced by Aharonov, Albert, and Vaidman[13–17]. This special measurement process involves three steps, quantum state preparation (pre-selection), a weak interaction, and post-selection on a final quantum state, which is nearly orthogonal to the initial state. The outcome of such measurement, the so-called weak value, may not only become exceedingly large and lie outside the eigenvalue spectrum of the observable but can also assume complex values[13–15]. The quantum mechanical concept of WVA can be understood within the realm of wave optics and thus most of the experiments on weak measurements are performed in classical optical setting [17–23].The WVA concept can be interpreted as near destructive interference between the eigenstates of the measuring observable as a consequence of nearly mutually orthogonal pre and post selections of the system states[13,14]. Based on this simple yet profound interferometric philosophy of WVA, we had initially shown that a similar situation arises in the spectral domain interference of Fano resonance that leads to natural WVA of small Faraday rotation and ellipticity in WMPC[12], which act as the weak interaction in this scenario. An analytical model for natural WVA in Fano resonance demonstrating the amplification of small polarization anisotropic effects around the Fano destructive point was given in our previous publication. However, in the context of the WMPC, such a model[12] was done for a specific case where the resonance frequencies for the TM and TE quasiguided modes are the same. Yet, it is important to investigate a more general scenario of natural WVA of Faraday effects in Fano interference, which will provide a better understanding on the role of WVA in the enhancement of MO properties in magneto-plasmonic crystals. Here, we have therefore addressed this by adopting differences in the resonance frequencies between the TE and TM waveguide modes. This helps us revealing the dependence of WVA of Faraday effects in Fano resonance (for TE excitation) on the resonance frequencies of the TM and TE modes and their varying overlap which can be tuned by changing the geometrical parameter of the plasmonic crystal. In contrast, for TM polarization excitation, the underlying physical mechanisms for the MO effect enhancement are expected to be rather complex and more interesting due to simultaneous involvement and coupling of the waveguide mode and the plasmon mode in such systems. As there appears to have different possible mechanisms of the enhancement of MO effects in magneto-plasmonic crystals[2–4,10] including the one of natural WVA[12], the question arises whether there is a common physical origin for the enhancement irrespective of the input polarization state or are there distinctly different mechanisms with varying strengths depending upon the excitation polarization and the geometrical parameters of the WMPC. The main purpose of this work is to address this important question and to shed light upon the competing roles of the different contributing mechanisms, namely, the natural weak value amplification and the purity of Fano resonance[12], near field enhancement effects due to the excitation of the surface plasmons[2], resonance enhanced cross coupling between the TE and TM polarization[3,4], the waveguide-plasmon strong coupling [3,4] and the related phenomenon of avoided crossing and so forth on the giant enhancement of magneto-optical effects (Faraday rotation and ellipticity).

The paper is organized as follows. We first provide the theoretical framework of natural interferometric WVA in Fano resonance and the results of the enhancement of Faraday rotation and ellipticity with TE polarization excitation in WMPC having varying geometrical parameters and the corresponding natural weak value interpretation are presented subsequently. We then present the simulation results for the enhancement of MO effects with TM polarization excitation for similar WMPCs with varying geometrical parameters. The intriguing roles of the electromagnetic near field enhancement due to the excitation of the surface plasmons, the waveguide-plasmon strong coupling and the avoided crossing behaviour[24,25] on the enhancement of the Faraday rotation and ellipticity are analyzed and interpreted. The paper concludes with a summary of the gained quantitative understanding on the different physical mechanisms of the MO effect enhancement, which in turn provides a useful recipe for tailoring their contributions to achieve optimal enhancement through optimization of the geometrical parameters of the nanostructure. The details of the WMPC structures and the specifics of the numerical simulations using Finite element method (FEM) are given in the Methods section.

## Results

### *Theoretical Framework of natural weak value amplification of Faraday effects in a Fano resonant WMPC*

We have recently developed a theoretical treatment for natural WVA of polarization rotation and ellipticity in the spectral domain interference of Fano resonance [12]. Briefly, in this model, the spectral ($\omega$) variation of the polarized electric field of Fano resonance[26,27] is modelled as the interference of a $y$-polarized frequency-independent continuum mode with an optically active narrow resonance mode exhibiting small rotation($\alpha$) and ellipticity($\chi$). However, in the context of the WMPC this WVA model[12] was demonstrating WVA of the Faraday effects for a specific case, where both TM and TE waveguide modes have the same resonance frequency providing the ideal scenario of natural WVA. Here we extend this model to a more generalized scenario of Fano interference in the WMPC by considering different resonance frequencies for the TE and the TM waveguide modes. The resultant polarized field can be expressed as

$$E_s(\omega) \approx= \left[\frac{q-i}{\varepsilon(\omega)^{TE}+i}\left(\alpha\ \cos\cos\chi\ - i\sin\sin\alpha\ \sin\sin\chi\right)\hat{y} + \frac{q-i}{\varepsilon(\omega)^{TM}+i}\left(\sin\sin\alpha\ \cos\cos\chi\ + i\cos\cos\alpha\ \sin\sin\chi\ \right)\hat{x} + \hat{y}\right] \quad (1)$$

Here, $\varepsilon(\omega)^{TE} = \frac{\omega-{\omega_0}^{TE}}{(\gamma/2)}$, $\varepsilon(\omega)^{TM} = \frac{\omega-{\omega_0}^{TM}}{(\gamma/2)}$, where ${\omega_0}^{TE}$ and ${\omega_0}^{TM}$ are the resonance frequencies of the TE and the TM waveguide modes respectively and $\gamma$ is the width of the resonance assumed to be similar. The $q$ parameter is related to the coupling of modes and determines the characteristic asymmetry of Fano spectral line shape. WVA of small polarization rotation $(\alpha)$ and ellipticity$(\chi)$ takes place as a consequence of near destructive spectral domain interference between the two modes with slightly different polarization states (which arises due to the presence of the small optical activity of the narrow resonance mode, acting as the weak interaction parameter in this weak measurement scenario) around the frequency of the Fano spectral dip $\omega_F = \left({\omega_0}^{TE} - \frac{q\gamma}{2}\right)$corresponding to $\varepsilon^{TE} = -q$. Note that $\omega_F$ is the frequency of exact destructive interference, where the phase difference between the TE quasiguided mode and incoming broad continuum becomes $\Psi(\omega_F) = \pi$ and the ratio of the amplitudes between them is unity $a(\omega_F) = 1$. At close vicinity of $\omega_F$, near destructive spectral interference takes place between the two modes with simultaneous small amplitude offset $\left(\epsilon_a(\omega)\right)$and phase offset $\left(\epsilon_p(\omega)\right)$. It is this near destructive spectral interference between the two modes that mimics the near orthogonal pre and post-selections of states (small overlap of states $\epsilon_{a/p}$) in conventional optical WVA. It is important to note that for the ideal case of natural WVA in Fano resonance ($\omega_0^{TE} = \omega_0^{TM} = \omega_0$), there is a common amplitude$(\epsilon_a)$ and phase offset $\left(\epsilon_p\right)$ parameter for both TE and TM waveguide modes, which are obtained in terms of the parameters of Fano resonance as

$$\epsilon_a(\omega) \approx \frac{\left(1-\sqrt{\left(\frac{\varepsilon^2+1}{q^2+1}\right)}\right)}{\left(1+\sqrt{\left(\frac{\varepsilon^2+1}{q^2+1}\right)}\right)};\ \epsilon_p(\omega) = \pi - \Psi(\omega) = \pi - tan^{-1}\left[\frac{q+\varepsilon}{1-q\varepsilon}\right] \qquad (2)$$

However, in the general case, the amplitude and phase offset parameters for TE $(\epsilon_{a/p})$ and TM $(\epsilon'_{a/p})$ interfering modes will differ owing to their different resonance frequencies, which can be expressed as

$$\epsilon_a(\omega) \approx \frac{\left(1-\sqrt{\left(\frac{\varepsilon^2+1}{q^2+1}\right)}\right)}{\left(1+\sqrt{\left(\frac{\varepsilon^2+1}{q^2+1}\right)}\right)};\ \epsilon_p(\omega) = \pi - \Psi(\omega)^{TE};\varepsilon(\omega) = \frac{\omega-{\omega_0}^{TE}}{(\gamma/2)}$$

Similarly

$$\epsilon_a'(\omega) \approx \frac{\left(1-\sqrt{\left(\frac{\varepsilon^2+1}{q^2+1}\right)}\right)}{\left(1+\sqrt{\left(\frac{\varepsilon^2+1}{q^2+1}\right)}\right)};\ \epsilon_p'(\omega) = \pi - \Psi(\omega)^{TM}; \varepsilon(\omega) = \frac{\omega-{\omega_0}^{TM}}{(\gamma/2)} \tag{3}$$

Using the similar framework of WVA[12] it can be shown that as in the case of ideal WVA, the real and the imaginary WVA of the polarization rotation $\alpha$ and ellipticity $\chi$ are manifested as $\epsilon_{a/p}$dependent large changes in the polarization state of light (Stokes vector elements), which acts as the pointer here (see Supplementary note 1), albeit with the two set of the small amplitude and the phase offset parameters ($\epsilon_{a/p}$for TE and $\epsilon_{a/p}'$ for TM) incorporated in the WVA equation (Eq. S6 of Supplementary information). One important difference is that, as the TE and TM waveguide modes spectrally moves away from each other, one cannot simultaneously obtain a very small value of both set of the amplitude and phase offset parameters $\epsilon_{a/p}$ and $\epsilon_{a/p}'$ at any given frequency around the Fano spectral dip $\omega_F$, which is desirable for optimal weak value amplification. Consequently, the maximum achievable amplification of polarization rotation $\alpha$ and ellipticity $\chi$ gets limited in this scenario. In contrast, for the ideal WVA scenario ($\omega_0^{TE} = \omega_0^{TM} = \omega_0$), this condition is easily obtained around $\omega_F$ leading to the maximum achievable amplification. We emphasize that the WVA does not break down even if one is observing the amplification at relatively larger amplitude and phase offsets, only the maximum magnitude of WVA goes down with slight deviation from the ideal behaviour, which is more accurate for small offsets (see also results of the model presented in Supplementary figure 1).

***Enhancement of MO effects in WMPCs for TE polarization excitation as a manifestation of interferometric natural weak value amplification in Fano resonance***

The WMPC system comprises of a periodic array of noble metal nanostructures on top of a dielectric waveguiding layer, which is made of MO materials that exhibit Faraday rotation and ellipticity. The coupling of the spectrally broad surface plasmon mode in the metallic nanostructures or the photon continuum (depending upon the excitation polarization of light) with the narrow quasiguided photonic modes in the waveguiding layer leads to Fano resonance. We used finite element method (FEM)[29] for simulating the polarization resolved transmitted spectra from such system (see Methods). A schematic illustration of the FEM simulation of transmission spectra and MO responses of the WMPC is presented in Figure 1. The WMPC system investigated in this study consists of gold (Au) grating on top of a thin yttrium-bismuth iron garnet (Y-BIG) film and the substrate is taken to be quartz. The MO active Y-BIG film serves as the waveguiding layer and additionally exhibit Faraday effect in the presence of an external magnetic field (see Methods). The simulated transmittance spectra $(E = \hbar\omega = 0.83\ to\ 2.067eV)$and Faraday rotation and ellipticity for varying geometrical parameters of the WMPC with TE polarization excitation are summarized in Figure 2. The transmission spectra (1st row), the spectral variation of the observed Faraday rotation $\psi$ (2nd row) and the observed net ellipticity $\xi$ (3rd row) of the WMPC are shown for varying periodicity (*d,* 1st column), width ($w$, 2nd column) and height ($h$, 3rd column) of the Au grating. Several observations are in place. Prominent signatures of Fano resonance with asymmetric spectral line shape are observed in all the transmission spectra with TE polarization excitation (1st row). Here, the surface plasmons are not directly excited, and hence the Fano resonance[30–32] arises due to the interference between the spectrally narrow quasiguided resonance mode with an ideal photon continuum. Importantly, giant enhancements of the Faraday rotation (2nd row) and ellipticity (3rd row) are observed around the Fano spectral dip ($E = \hbar\omega_F$) for all the geometrical parameters of the WMPC. In agreement with previous reports[25], the Fano dip is observed to gradually move towards shorter frequency (or energy $E = \hbar\omega$) with increasing periodicity (*d*) of the grating (Fig. 2a). The magnitudes of the enhanced Faraday rotation $\psi$ (Fig. 2b) and ellipticity $\xi$ (Fig. 2c) exhibit an interesting trend with varying *d*, which increase to a maximum value for a certain value of periodicity and then follows a gradual decrease. The maximum magnitudes of the enhanced Faraday rotation ($\psi \sim 1.551°$) and ellipticity ($\xi$ ~$1.559°$) of the WMPC are obtained for a value of *d*= 550 nm (Fig. 2b), as compared to the corresponding average bare film rotation and ellipticity of $0.28°$ and $0.1°$, respectively. Neither the value of the width *w* nor the height $h$ of the Au grating has any significant influence on the quasiguided modes and hence the resulting Fano spectral line shape of the WMPC is not sensitive to variations of the *w* and $h$-parameters of the grating (Fig. 2e, 2i). Accordingly, the magnitude and spectral variation of both the Faraday rotation (Fig. 2j) and ellipticity (Fig. 2k) do not exhibit appreciable variation with the change in the height $h$ of the grating. Similarly, with varying width *w* of the gating, the Faraday rotation $\psi$ (Fig. 2f) and ellipticity $\xi$ (Fig. 2g) exhibit only very weak variations (Faraday effects seem to be marginally stronger for *w* = 120 nm as compared to others). The spectral variation of the transmitted TM polarization intensities with TE polarization excitation are also plotted for each of the varying geometrical parameters of the WMPC (Fig.2 4th row). The overall TM polarized intensities are considerably weak and approach their respective minima near the Fano spectral dip where maximum magnitudes of Faraday rotation and ellipticity are observed. This rule out the possibility of the role of the resonance enhanced cross-coupling of the TE and the TM quasiguided modes in the enhancement of the Faraday effect. The observed enhancement of the Faraday rotation and ellipticity around the Fano spectral dip corresponding to near destructive spectral domain interference between the quasiguided mode and the photon continuum thus points towards the vital role of the natural WVA of Fano resonance for TE polarization excitation. This aspect is therefore investigated in further details by studying the dependence of the Faraday effect enhancement on the periodicity of the grating and its natural weak value interpretation. We have also provided the near field distribution of both TE and TM around the Fano dip (Supplementary figure 3) which confirms the dominance of TE polarized field and the waveguiding nature of both TE and TM around the Fano dip.

Figure 3 summarizes the results of natural WVA interpretation of Faraday rotation and ellipticity enhancements for TE polarized excitation for Fano resonant WMPCs with two different values of periodicity of the Au grating ($d$ = 550 nm, Fig. 3A and $d$= 650 nm Fig. 3B) and having the same width and height $w$= 120 nm and $h$ = 100 nm, respectively. The spectral $(E = 0.83\ to\ 2.067eV)$ variations of both the Faraday rotation $\psi$ (Fig. 3A(a) and Fig. 3B(a)) and ellipticity $\xi$ (Fig. 3A(b) and Fig. 3B(b)) are plotted along with the transmitted intensity profiles of the WMPCs. In order to analyze these results using the natural WVA formalism of Fano resonance, the spectral line shape of the transmitted intensity is first fitted with Fano intensity formulae (Eq. S3 of Supplementary information). The relevant Fano resonance parameters, namely, the central frequency of the interfering narrow resonance of the quasiguided mode ${\omega_0}^{TE}$,width of the narrow resonance $\gamma$, the Fano spectral asymmetry parameter $q$ are estimated from this fitting. These are subsequently used to determine the exact values of the frequency corresponding to the Fano dip $\omega_F = \left( {\omega_0}^{TE} - \frac{q\gamma}{2}\right)$and the amplitude offset ($\epsilon_a$) and the phase offset($\epsilon_p$) parameters (using Eq. 2) in the natural WVA formalism. In scenario, where the resonance frequencies of the TM and TE quasiguided modes perfectly overlaps $(\omega_0^{TE} = \omega_0^{TM} = \omega_0)$there is a common amplitude$(\epsilon_a)$ and phase offset $\left(\epsilon_p\right)$ parameter (as per Eq. 2 and Eq. 3). As discussed previously at the frequency corresponding to the spectral domain destructive interference $(\omega_F)$, both the offset parameters $(\epsilon_{a,}\epsilon_p)$ are zero and as one moves spectrally slightly away from $\omega_F$, both the $\epsilon_{a,}\epsilon_p$ weak measurement parameters vary simultaneously (shown in Fig. 3A (d)). The complex (simultaneous real and imaginary) WVA of the weak Faraday effect is accordingly manifested as large enhancement of both the Faraday rotation and ellipticity when$(\epsilon_{a,}\epsilon_p \to 0)$ for the WMPC with $d$ = 550 nm (Fig. 3A (e) and (f)). In order to further comprehend the important role of the natural WVA of Fano resonance, we have shown the comparison of the transmitted TM and TE polarized intensities with TE polarized excitation (Fig. 3A (c)). This clearly shows that the transmitted TM polarized intensity is much weaker than the TE polarization signal and both approach their respective minima in the vicinity of the Fano spectral dip where the rotation reaches its maximum value. This indicates that the amplification of the weak Faraday effect arises due to the near destructive interference of the TE polarized quasiguided mode with the continuum mode at close vicinity of the Fano spectral dip and it takes place at the expense of the total intensity signal, which is a universal signature of WVA. The simulated Faraday rotation $\psi$ and ellipticity $\xi$ for the WMPCS having $d$= 550 nm also follows the $\left(\propto \alpha cot\, \epsilon_{a/p} \sim \frac{\alpha}{\epsilon_{a/p}}\right)$ behaviour (Fig. 3A (e) and (f)) as predicted by the natural WVA formalism. A quantitative comparison of the amplified Faraday rotation $\psi$ (Fig. 3A (e)) and the ellipticity $\xi$(Fig. 3A (f)) with the exact theoretical predictions of natural WVA of Faraday effect in Fano resonance (using Eq. S4 and S6 of Supplementary information) shows excellent agreement. This confirms that natural WVA in Fano resonance is the underlying reason for the enhancement of weak Faraday effect in the WMPC system having a value of periodicity $d$ = 550 nm. We now move on to inspect similar results for the WMPC with periodicity $d$ = 650 nm (Fig. 3B). An important difference between the WMPC with $d$ = 550nm as compared to that with $d$ =650nm is that while for the latter the resonance frequencies of the TM and TE quasiguided modes perfectly overlaps $(\omega_0^{TE} = \omega_0^{TM})$, for the former these two frequencies are separated $(\omega_0^{TE} \neq \omega_0^{TM})$. Consequently, the amplitude and the phase offset weak value parameters $\epsilon_{a/p}$for TE and $\epsilon'_{a/p}$for TM interfering modes will differ owing to their different resonance frequency, and the amplification of Faraday effect will explicitly depend on both set of the amplitude and phase offset parameters, $\epsilon_{a/p}$and $\epsilon'_{a/p}$. These offset parameters corresponding to the TE and TM waveguide modes are shown in Fig. 3B (c) and (d) respectively. From the presented results (Fig. 3B (c) and (d)) it is evident that both set of the offset parameters $\epsilon_{a/p}$ and $\epsilon'_{a/p}$do not simultaneously approach zero at any frequency around the Fano spectral dip $\omega_F$, which is desirable for optimal weak value amplification of Faraday effect. Hence the maximum achievable amplification of both the Faraday rotation and ellipticity get limited with increasing difference between $\omega_0^{TE}$ and $\omega_0^{TM}$. In contrast, when the resonance frequencies are same for both TE and TM waveguide modes$(\omega_0^{TE} = \omega_0^{TM} = \omega_0)$, one obtains the ideal weak value amplification scenario around the Fano spectral dip leading to maximum enhancement of Faraday effects. This indeed is the case as the enhancement of both the Faraday rotation (Fig. 3B(a)) and ellipticity (Fig. 3B(b)) are considerably weaker for the WMPCS with $d = 650$nm as compared to that for $d = 550$ nm (Fig. 3A(a) and Fig. 3A(b)). In such case also, even though the exact theoretical predictions of natural WVA of Faraday effect in Fano resonance marginally deviates from the observed Faraday rotation$\psi$ (Fig. 3B (e)) and ellipticity$\xi$ (Fig. 3B (f)), it still provides reasonable agreement.

To summarize, the results presented in Fig.2 and Fig. 3 provide conclusive evidence that natural WVA of Faraday effect in the spectral domain interference of Fano resonance is the primary mechanisms for the enhancement of Faraday rotation and ellipticity in WMPCs with TE polarization excitation. Moreover, strong spectral overlap between the TM ($x$) and TE ($y$) quasiguided resonance modes provides an ideal scenario for natural WVA in Fano resonance which leads to achieve an optimal enhancement of Faraday rotation and ellipticity in the WMPCs. It is also important to note that the WVA approximation does not break down even if the resonance modes are spectral away from each other, only the amplification value goes down with possibly slight deviation from the ideal WVA behaviour which is known to be more accurate for small magnitudes of the amplitude and the phase offset parameters. Therefore, it may be logical to conclude that in this scenario the same mechanism of natural WVA works even if the resonance frequencies of the two modes are not the same, albeit with weaker enhancement effect.

It is pertinent to note here that, the amplification of any weak effect within the framework of WVA always happen at the cost of the total intensity signal. Despite loss of the signal, it has been well demonstrated[15-17,19] that WVA leads to significant enhancement of actual signal to noise ratio as it throws away redundant signal and picks up only the relevant part of the signal responsible for the amplification effect. The enhancement of Faraday rotation in our WMPC system with TE polarization excitation is no exception in this regard. Like other usual optimization strategy of WVA, here also optimal amplification of Faraday effects and moderate transmittance can be obtained simultaneously by carefully choosing a frequency window around the Fano spectral dip (see Supplementary figure 2 for details). Note that for photonic device applications, it is desirable to have simultaneous enhancement of both the optical transmission and the Faraday rotation angle. In order to evaluate the practical performance of a magneto-optic device, the magneto-optical figure of merit (MOFOM) is usually defined as $MOFOM = \left(|\theta| \times \sqrt{T}\right)$, where $\theta$ is the Faraday rotation angle and $T$ is the transmittance coefficient[4]. In the inset of Figure 3A (a), we have therefore shown the spectral variation of the MOFOM of the WMPC system around the Fano spectral dip (in the region of interest) and compared it with that of the bare film. This figure clearly illustrates the advantage of using the WMPC system and provides strong evidence of the WMPC system being a potential candidate for building nonreciprocal photonic devices. It is also important to note in this regard that in all the actual optical Fano resonant systems (like the WMPC), the scattered intensity never reaches zero at the Fano spectral dip due to the presence of finite dissipation in practical systems. This may also therefore enable one to work even at the exact Fano dip, where the maximum amplification can be utilized. We now turn our attention to the enhancement of Faraday effect in WMPCs with TM polarization excitation, where the surface plasmons are directly excited and hence the underlying mechanism is expected to be fundamentally different and more complex.

***Enhancement of MO effects in WMPCs for TM polarization excitation: role of the waveguide-plasmon strong coupling and avoided crossing***

With TM polarization excitation, the spectrally narrow quasiguided mode and the broad surface plasmon modes are simultaneously excited, leading to hybridization and strong coupling of waveguide-plasmon modes. As a result of this, the waveguide-plasmon hybrid mode splits the resonance and two dips are observed in the transmission spectra ($E = 0.83\ to\ 2.067eV$)of the WMPC (Figure 4, 1st row). Even though there is mild spectral asymmetry in the transmission spectra, the Fano resonance is not as prominent as that for TE polarization excitation. One reason for this is the fact that in case of TE polarization excitation, the photon continuum provides an ideal continuum as desirable for ideal Fano resonance, whereas the broad surface plasmon mode has finite line width and spectral profile. Despite the fact that one has hybridized waveguide-plasmon modes, the dominant contributions of the two different modes in the observed spectral dips can still be distinguished for certain geometrical parameter range of the WMPC. For example, at lower value of the periodicity of the grating $d$ = 400nm, the transmission dip at the lower frequency end (energy $E = \hbar\omega \sim 1.38\ eV$)shows broader spectral feature which therefore appears to be dominated by the broad surface plasmon mode (Fig. 4a). The sharp waveguide resonance, on the other hand, dominates the behaviour of the narrower dip at the higher frequency end ($E \sim 1.91eV$) (discussed subsequently). With increasing value of the periodicity $d$ of the grating, the two modes (waveguide mode and the plasmon resonance mode) approach spectrally closer to each other and a clear avoided or anti-crossing behaviour[24,25] is observed as a universal manifestation of the strong coupling phenomenon (Fig. 4a). It is however, important to note that large enhancement of both the Faraday rotation $\psi$ (Fig. 4b) and ellipticity $\xi$ (Fig. 4c) are always observed in the spectral dip at the lower frequency end which is primarily dominated by the surface plasmon mode for lower value of the grating periodicity (below the avoided crossing regime) and is dominated by the waveguide mode at higher grating periodicity (above the avoided crossing).Thus, both the electromagnetic near field enhancement associated with the surface plasmons and the formation of the strongly coupled waveguide-plasmon polariton system play important role in this enhancement. A trend similar to that observed for the TE excitation is also observed here with varying periodicity *d* of the grating (Fig. 4, 1st column), the magnitudes of the enhanced Faraday rotation $\psi$ (Fig. 4b) and ellipticity $\xi$ (Fig. 4c) attain a maximum value for a specific value of $d$ (500 nm) and then decreases. Here, the maximum enhancement of the rotation ($\psi$~1.536°) and ellipticity ($\xi$~2.68°) are obtained for the range of *d* ~475– 550 nm. This is particularly interesting as this periodicity range is exactly within the window of the avoided crossing regime where the splitting is minimum and the two modes are spectrally the closest (shown subsequently). This underscores the vital role of the strong coupling and the avoided crossing of the waveguide-plasmon modes in the resulting enhancement of Faraday effect. Note that a very similar trend is also observed for varying width *w* of the grating in the WMPC (Fig. 4 2nd column). The avoided crossing behaviour of the two modes appears in the transmission spectra since the width parameter *w* changes the spectral position of the plasmon resonance (Fig. 4e). The maximum enhancement of the Faraday rotation (Fig. 4f) and ellipticity (Fig. 4g) takes place for 120nm grating width, which is once again within the window of the avoided crossing regime of minimum splitting. No significant changes in the transmission spectra (Fig. 4i) are observed for the variation of height $h$ of the grating as it neither affects the spectral response of the plasmons or the waveguide resonances (Fig.4 3rd column). For each of the cases of varying periodicity, width and the height of the grating in the WMPCS, the transmitted TE polarization spectra are also shown with TM polarization excitation (Fig. 4, 4th row).

In order to further comprehend the role, the avoided crossing behaviour associated with the strong coupling of the waveguide-plasmon modes, the frequencies (energy $E = \hbar\omega$) corresponding to the two spectral dips are shown as a function of varying value of periodicity *d* of the grating in the WMPC (Fig. 5a). A clear avoided crossing behaviour is apparent and the minimum spectral difference (minimum splitting) between the two branches appears for the grating periodicity $d$~ $475 - 550$ nm, for which the maximum enhancement of the magnitudes of the Faraday rotation $\psi$ (Fig. 5b) and ellipticity $\xi$ (Fig. 5c) are observed. To understand the role of the hybridization of the plasmon and the waveguide modes, the simulated spatial distribution of the magnetic field for input TM polarization ($H_y$) at the energies (frequencies) corresponding to the two transmission dips are shown for three different grating periods. The first one is for grating period $d$ =400 nm (Fig. 5d) below the avoided crossing regime, the second one is for $d$ =500 nm (Fig. 5e) which is exactly at the avoided crossing regime corresponding to minimum splitting, and the 3$^{rd}$one is for $d$ =650nm (Fig. 5f) above the avoided crossing. For grating period $d$ =400nm below the avoided crossing, the field distribution corresponding to the spectral dip at the lower frequency ($E$ = 1.38 eV) shows dominant character of plasmon resonance, whereas the waveguide field distribution is more visible at the longer frequency ($E$ = 1.91eV) region (Fig. 5d). Above the avoided crossing of the modes, the natures of the two modes flip which can be clearly observed from the spatial field distribution for $d$ =650nm grating period (Fig. 5f). The lower frequency ($E$ =1.14 eV) dip now shows prominent signature of waveguide field distribution and the dip at the higher frequency ($E$ =1.46eV) is dominated by the plasmon mode. For grating period $d$ =500nm near the avoided crossing, a stronger hybridization of the waveguide-plasmon mode is observed, where maximum enhancement of the rotation and ellipticity are also observed. Note that the enhancement of Faraday effect is observed in the lower frequency spectral dip in the avoided crossing regime, where the electromagnetic field enhancement is maximum (Fig. 5d, e, f). It is also extremely crucial in this regard that in the window of the avoided crossing (at ~ $d$ =500 nm grating periodicity), the plasmon and the waveguide modes are spectrally nearest to each other leading to the intersection of the TE waveguide mode dispersion curve with the lower frequency dip of the TM-excited hybrid waveguide-plasmon mode (shown in Fig. 5a). This therefore provides an ideal scenario of resonance enhanced cross-coupling between the TM-TE polarizations enabling maximum enhancement of Faraday effect in this region.

## Discussion

In summary, our studies have unravelled distinctly different physical origins for the enhancement of Faraday rotation and ellipticity in magneto-plasmonic crystals with TE and TM polarization excitation. Natural weak value amplification of weak Faraday effect that arises due to near destructive spectral domain interference in Fano resonance of the plasmonic crystals is identified as the primary mechanisms for enhancement with TE polarization excitation. The results also showed that the enhancement of the magneto-optical effects are maximum for those geometrical parameters of plasmonic crystals which exhibit a strong spectral overlap of the TE and TM quasiguided modes, in which case an ideal weak value amplification takes place. In contrast for TM polarization excitation, electromagnetic near field enhancement associated with the surface plasmons and the strong coupling of the waveguide-plasmon mode plays the dominant role in the Faraday effect enhancement. The enhancements are the strongest near the avoided crossing regime where the waveguide-plasmon modes are spectrally the closest and the splitting is minimum. These results are of both fundamental and applied interests. The quantitative understanding gained on the different physical mechanisms for the enhancement of the magneto-optical effects provides a useful recipe and a systematic approach for the optimization of the geometrical parameters of the WMPC towards achieving maximum enhancement of Faraday effects, and for controlled tailoring of the magneto-optical responses of such hybrid systems. The distinctly different mechanisms for the enhancement of magneto optical effects for TE and TE polarizations also indicate that the choice of input polarization state may provide an extra degree of freedom to optimally combine the two different mechanisms for the enhancement of magneto-optical response of such hybrid magneto-plasmonic systems. These may altogether lead to optimized development of multifunctional non-reciprocal photonic nano-devices and potentially enhance their resulting applications in optical isolation, modulation, rotation, magnetic field sensing and so on. We also envisage that the understanding of the underlying mechanisms may also open up interesting direction of research in this promising area. For example, the observed enhanced Faraday effects in the absence of plasmon resonances may open up a new avenue where the newly understood 'near destructive interferometric origin' of enhancement can be used to observe significant enhancement of Faraday effects in all-dielectric nanostructures. This may be of particular interest as unlike the conventional metal-based magneto-plasmonic systems, these all dielectric hybrid magneto-optic systems are expected to have much lesser loss yielding high quality factor resonances and thus potentially enhancing ensuing applications. Moreover, the interferometric origin of natural weak value amplification of Faraday effect also opens up the possibility of extending the philosophy to some other intriguing wave phenomena like electromagnetically induced transparency (EIT) and absorption (EIA), extraordinary optical transmission (EOT), which have common origin in fine interference effects.  We are currently expanding our investigations in the aforementioned directions.

## Methods

### *Finite element Method (FEM) simulation of the WMPCs*

A schematic illustration of the FEM simulation[29] of transmission spectra and MO responses of the WMPC is presented in Figure 1**.** The WMPC system consists Au grating on top of a thin Y-BIG film with quartz substrate. The MO active Y-BIG film is the waveguiding layer and exhibit Faraday effect in the presence of an external magnetic field. For input TE ($y$) polarization, the electric field of light is oriented parallel to the axis of the Au grating, whereas for TM ($x$) polarization, the electric field is perpendicular to the grating (Fig. 1a). All the simulations are performed for normal incidence of light and the thickness of the waveguide layer are kept constant ($t$ = 150 nm) (Fig. 1b). The far field optical transmittance spectra ($E = \hbar\omega = 0.83 \; to \; 2.067 eV$, corresponding to $\lambda = 1495.78 \; to \; 600.62$ nm) and the corresponding spectral dependence of Faraday rotation and ellipticity of the WMPC system are studied by systematically varying the grating periodicity ($d$), width ($w$) and height ($h$). For performing the simulation we have used a tensorial polarizability or a dielectric tensor which has the form $\epsilon = \left(\epsilon_{xx} = \epsilon_{11} \;\; -ig \;\; 0 \;\; ig \;\; \epsilon_{yy} = \epsilon_{33} \;\; 0 \;\; 0 \;\; 0 \;\; \epsilon_{zz} = \epsilon_{22} \right)$. Where the magnetic field induced off-diagonal elements of $\epsilon$ are the characterizing properties of the magneto-optic material (Y-BIG) and the origin for Faraday effects. The permittivity tensor elements of the Y-BIG film were taken, $\epsilon_{11} = \epsilon_{33} = 6.7 + 0.053i$ and $g = 0.016 - 0.0092i$ for typical magnetic field of 140mT[3] and dielectric permittivity of Au was taken from literature[33]. For performing the FEM simulation, the commercial software named COMSOL is used and consequently the conventional sign notation of the dielectric permittivity is used ($\epsilon_{11} = \epsilon_{33} = 6.7 - 0.053i$ and $g = 0.016 + 0.0092i$).

The spectral variations of the Faraday rotation and ellipticity from the transmitted intensity spectra of the WMPC were generated using standard Stokes polarization method[28]. Briefly, for a given input polarization of light (TM-$x$ or TE-$y$), six different polarization resolved intensity components of the transmitted light were recorded, namely, $I_H$– horizontal ($x$) linear polarization, $I_V$ – vertical ($y$) linear polarization, $I_P$ –45° linear polarization; $I_M$ – 135° linear polarization; $I_R$ – right circular polarization, $I_L$– left circular polarization. These intensities were used to yield the spectral variation of the Stokes parameters $[I, Q, U, V]^T$[28]. The observed net Faraday rotation (defined as polarization vector orientation angle $\psi$) and net Faraday ellipticity (defined as $\xi$) are determined from these Stokes parameters of transmitted light using Eq. 3a and Eq. 3c. Note that the angle $\psi$ is always defined with respect to the input polarization ($x$ or $y$).

**Data Availability:**

The datasets generated during and/or analyzed during the current study are available from the corresponding author on reasonable request.

**Code Availability:** The computer code or algorithm used to generate the data presented in the manuscript are available from the corresponding author on reasonable request.

**Competing Interests Statement:** The authors declare no competing interest regarding this work.

**Acknowledgement:** The authors acknowledge Indian Institute of Science Education and Research (IISER) Kolkata for the funding and facilities. S.G. acknowledges financial support from the Council of Scientific and Industrial Research (CSIR), Government of India through a research fellowship.

**Author Contributions:**

JKN and SG contributed equally to this work. SG, JKN performed all the FEM based numerical simulations of polarization resolved transmission spectra from the WMPCs, analyzed data, contributed in theoretical modelling and in writing the manuscript. AKS contributed in theoretical modelling and in data analysis. NG designed and supervised this study, contributed in theoretical modelling, analysis / interpretation of the results and in writing the manuscript.

**Figure Captions**

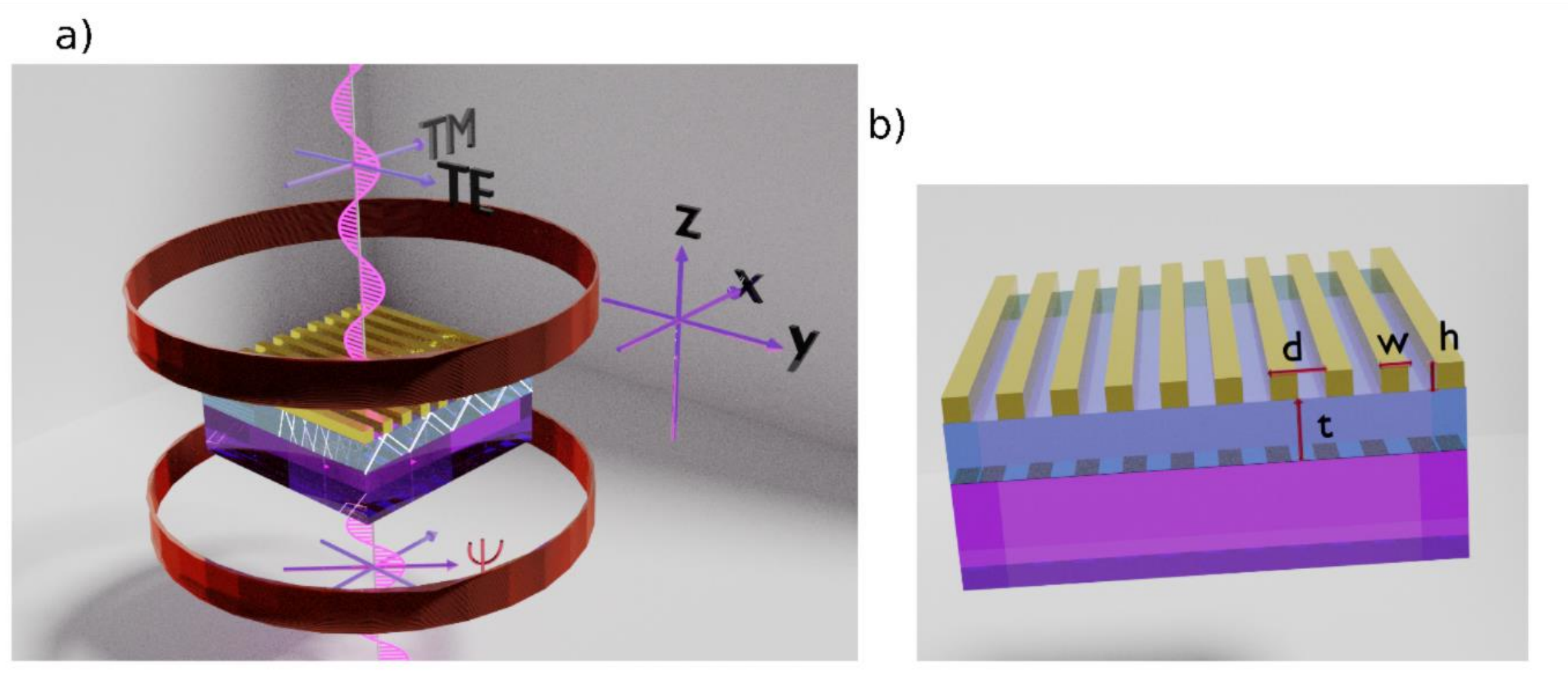

**Figure 1:** *The waveguide magneto-plasmonic crystal.* **(a)** A schematic illustration of the Finite element method (FEM) simulation of Faraday rotation $\psi$ in the transmission spectra of the WMPC system. The direction of the transverse magnetic TM (*x)* and transverse electric TE (*y*) polarized light electric fields are perpendicular and parallel (respectively) to the axis of the Au gratings in the WMPC. **(b)** The WMPC system comprises of gold (Au) gratings on top of a thin yttrium-bismuth iron garnet (Y-BIG) film and the substrate is taken to be quartz. The Y-BIG film serves as the waveguiding layer and additionally exhibit Faraday effect in the presence of an external magnetic field. The thickness of the Y-BIG film (*t*) and the periodicity (*d*), width (*w*), and height (*h*) of the Au gratings are marked.

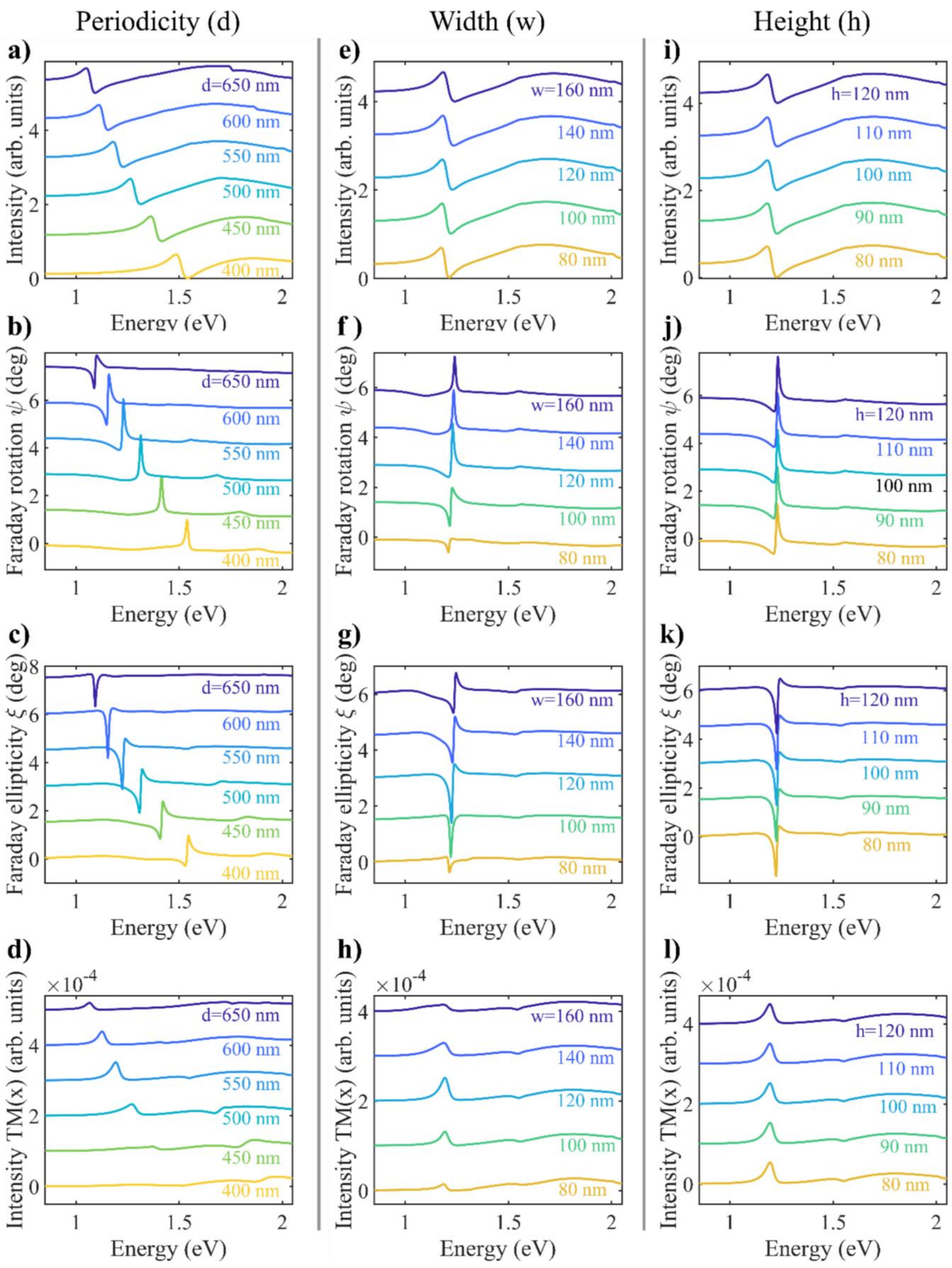

**Figure 2:** *The dependence of the enhanced Faraday rotation and ellipticity on the geometrical parameters of the Fano resonant WMPC system with TE polarized excitation.* The FEM simulated transmission spectra ((a), (e), (i)), the spectral variation of the Faraday rotation $\psi$ ((b), (f), (j)) and ellipticity $\xi$ ((c), (g), (k)) of the WMPC are shown for varying periodicity

($d$, (a), (b), (c)), width ($w$, (e), (f), (g)) and height ($h$, (i), (j), (k)) of the Au gratings (shown for $E = \hbar\omega = 0.83\ to\ 2.067eV$). Prominent signatures of Fano resonance with asymmetric spectral line shape are observed in all the transmission spectra ((a), (e), (i)). Large enhancements of the Faraday rotation ((b), (f), (j)) and ellipticity ((c), (g), (k)) are observed around the Fano spectral dip ($E = \hbar\omega_F$) for all the geometrical parameters. The spectral variation of the transmitted TM polarization intensities ((d), (h), (l)) is also shown for each of the varying geometrical parameters of the WMPC. The *d* parameter of the grating is varied ((a), (b), (c), (d)) from 400nm to 650nm with a step size of 50nm, keeping fixed *w= 120 nm* and *h= 100 nm.* Similarly, the *w* parameter of the grating is varied from 80 nm to 160 nm in a step of 20nm with a fixed *d*= 550nm and *h*= 100 nm. The *h* parameter is increased from 80nm to 120nm with a step of 10nm for a fixed *d*=550nm and *w*=120nm. In order to maintain clarity in the figures, the baseline magnitudes of Faraday rotation ψ and ellipticity ξ are shifted by ~1.5 deg. consecutively for each sub-plot with varying periodicity, width and height of the Au gratings.

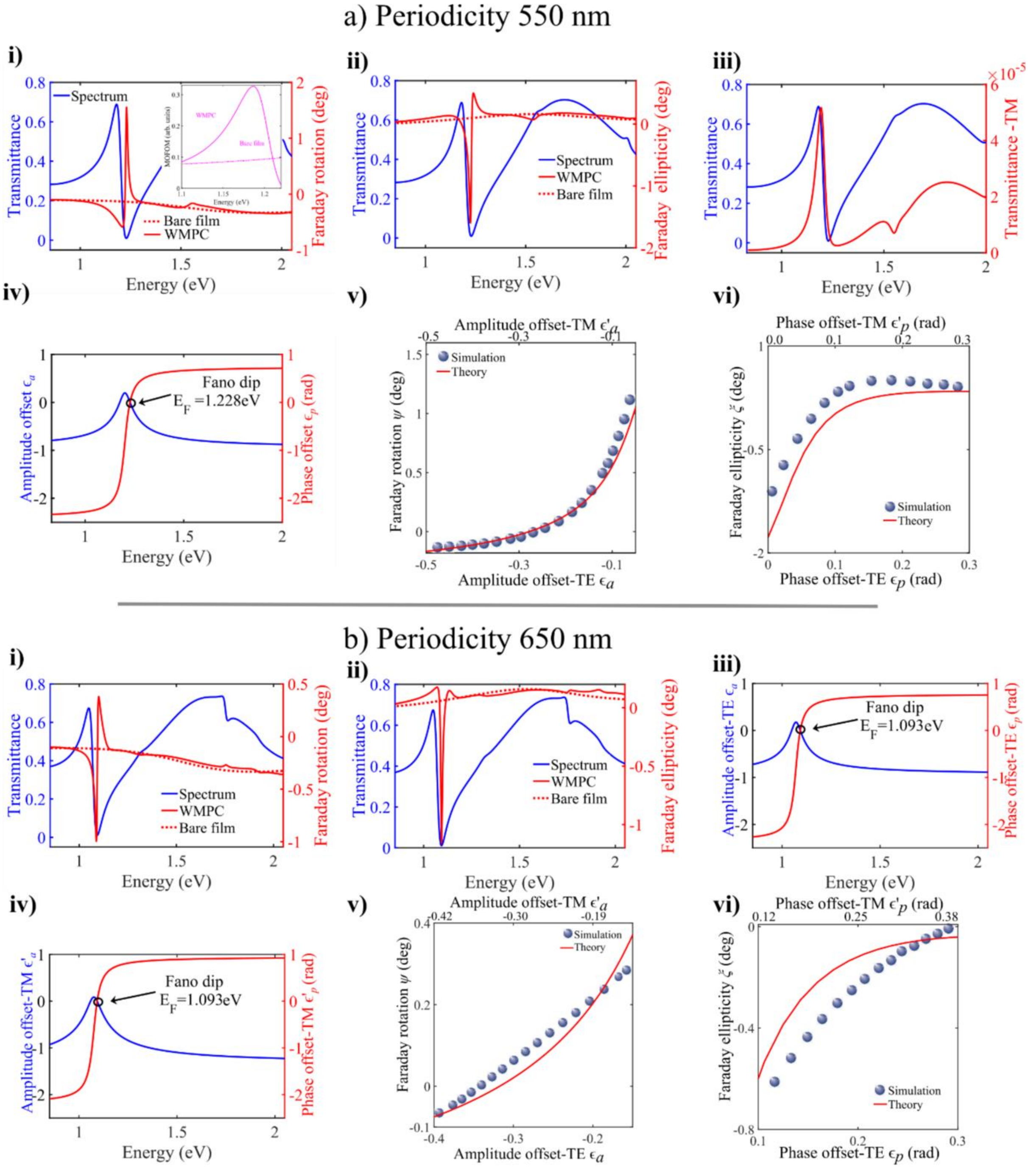

**Figure 3:** *Interpretation of enhanced Faraday rotation and ellipticity of WMPC for TE polarized excitation through natural WVA of Faraday effect in Fano resonance.* The simulation results are shown for two different periodicities $d$ of the grating in the WMPC: **(A)** $d$ = 550 nm and **(B)** $d$ = 650 nm. The spectral variations($E = \hbar\omega = 0.83\ to\ 2.067eV$)of Faraday rotation $\psi$(right axis, red solid line) (A**(a) and B(a)**) and ellipticity $\xi$ (right axis, red solid line) (**A(b) and B(b)**) are shown along with the transmitted intensity profiles (left axis blue solid line) and the bare film Faraday rotation (left axis red dotted line). In the inset of **Fig. A(a)**, the spectral

variation of the magneto-optical figure of merit (MOFOM) of the WMPC system (pink solid line) is shown around the Fano spectral dip region and compared with that of the bare film (pink dotted line). **(A(c)):** A comparison of the transmitted TM (right axis red solid line) and TE polarized intensities (left axis blue solid line) for TE polarized excitation**. (A(d), B(c) and B(d)):** Spectral variations of the weak measurement parameters, amplitude offset $\epsilon_a$(left axis, blue solid line) and the phase offset $\epsilon_p$(right axis, red solid line) around the Fano spectral dip ($E = \hbar\omega_F$) for both the waveguide modes. **(A(e) and B (e)):** The variation of the Faraday rotation $\psi$ as a function of amplitude offset parameters $\epsilon_a$ of TE and $\epsilon_a'$ of TM waveguide modes (grey metallic balls) and the corresponding theoretical predictions of natural WVA (red solid line) (using **Eq. S4- Eq. S6 of supplementary information**). The $\epsilon_a$and $\epsilon_a'$parameters are shown in the lower and upper $X$-axes, respectively. **(A(f) and B (f)):** The variation of the ellipticity$\xi$ as a function of the phase offset parameters $\epsilon_p$ of TE and $\epsilon_p'$ of TM waveguide modes (grey metallic balls) and the corresponding theoretical predictions of natural WVA (red solid line) (**using Eq. S4-S6 of Supplementary information**). The $\epsilon_p$and $\epsilon_p'$parameters are shown in the lower and upper $X$-axes, respectively.

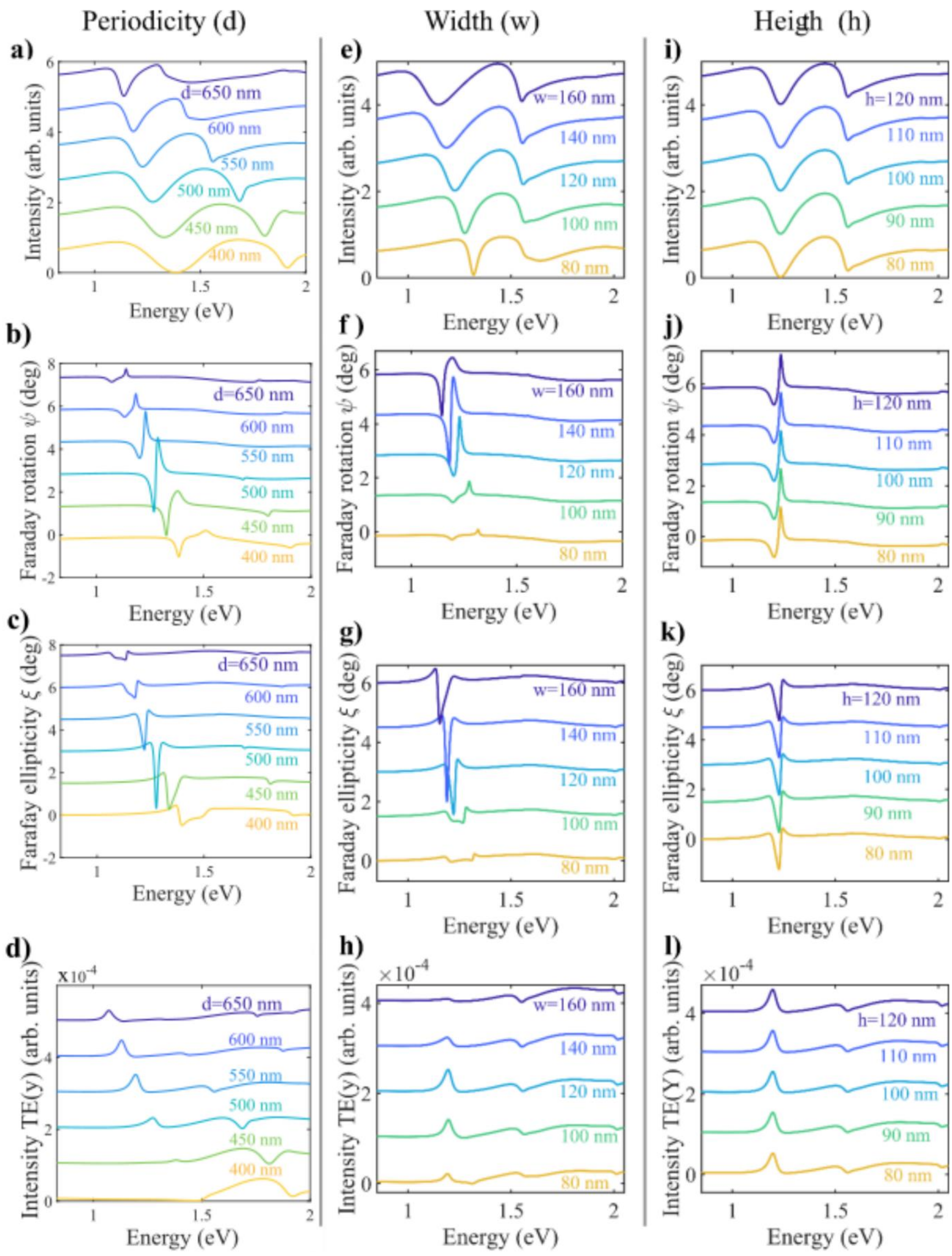

**Figure 4:** *The dependence of the enhanced Faraday rotation and ellipticity on the geometrical parameters of the WMPC system with TM polarized excitation.* The FEM simulated transmission spectra ((a), (e), (i)), the spectral variation of the Faraday rotation $\psi$ ((b), (f), (j)) and ellipticity $\xi$ ((c), (g), (k)) of the WMPC are shown for varying periodicity (*d,* (a), (b), (c)),

width ($w$, (e), (f), (g)) and height ($h$, (i), (j), (k)) of the Au gratings (shown for $E = \hbar\omega = 0.83\ to\ 2.067 eV$). Simultaneous excitation of waveguide and plasmon leads to the formation of hybrid plasmon-waveguide modes and consequently two dips appear in the transmission spectra (*1st row*). Large enhancements of the Faraday rotation (2nd row) and ellipticity (3rd row) are observed around the transmission dip at the lower frequency end. The spectral variation of the transmitted TE polarization intensities (4th row) are also shown for each of the varying geometrical parameters of the WMPC. The $d$ parameter of the grating is varied (1st column) from 400nm to 650nm with a step size of 50nm, keeping fixed $w$= *120 nm* and $h$= *100 nm.* Similarly, the $w$ parameter of the grating is varied from 80 nm to 160 nm in a step of 20nm with a fixed $d$= 550nm and $h$=100 nm. The $h$ parameter is increased from 80nm to 120nm with a step of 10nm for a fixed $d$=550nm and $w$=120nm. In order to maintain clarity in the figures, the baselines of each of the spectra are shifted. In order to maintain clarity in the figures, the baseline magnitudes of Faraday rotation $\psi$ and ellipticity $\xi$ are shifted by ~1.5deg. consecutively for each sub-plot with varying periodicity, width and height of the Au gratings.

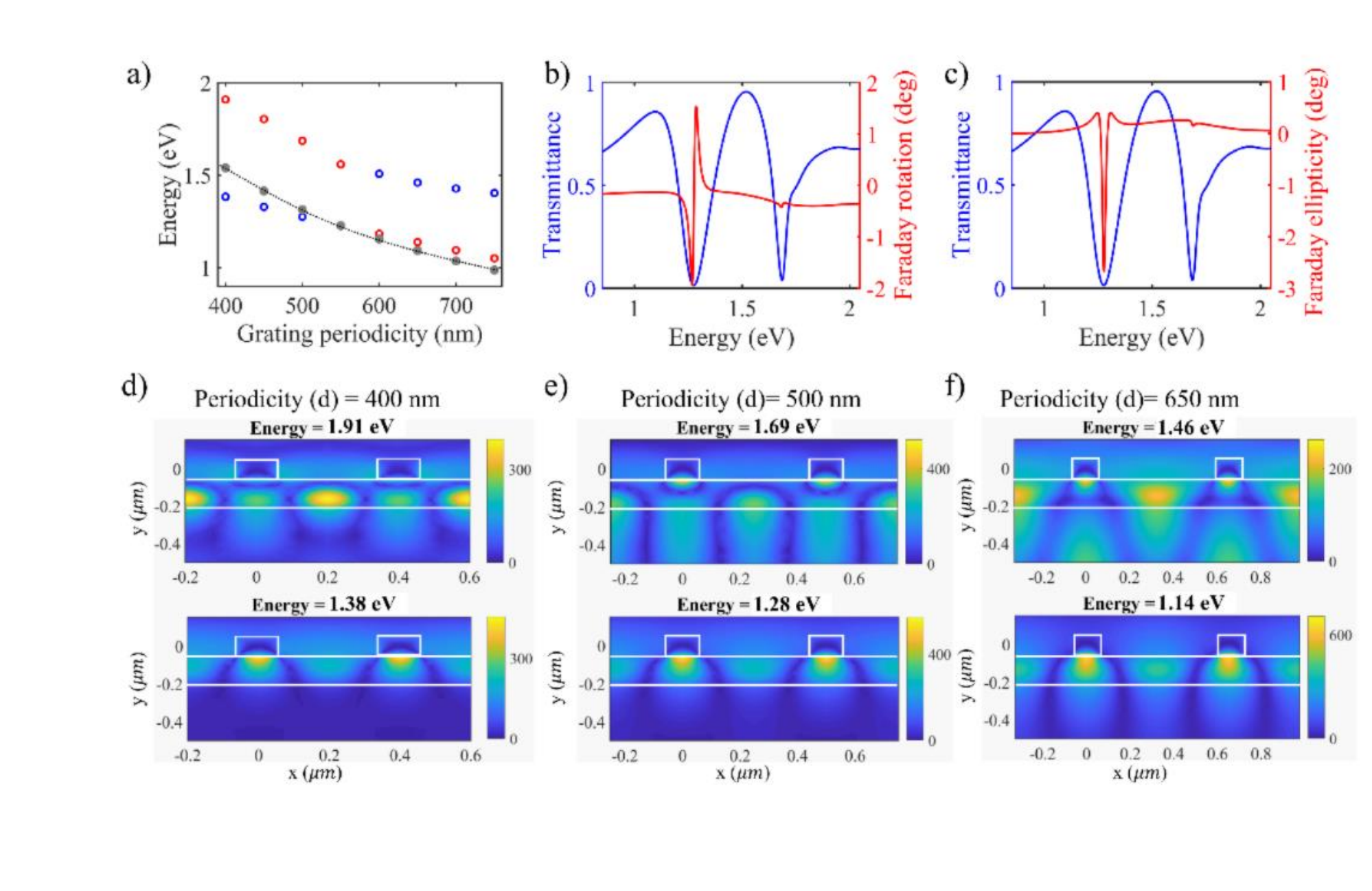

**Figure 5:** *Understanding the role of the waveguide-plasmon strong coupling and avoided crossing on enhanced Faraday effect in WMPC system with TM polarized excitation.* **(a)** The frequencies (energy $E = \hbar\omega$) corresponding to the spectral dips are shown as a function of varying value of periodicity $d$ of the grating in the WMPC width and height fixed at 120nm and 100nm respectively. The dispersion of the TE waveguide mode (black dotted line) is also shown for varying periodicities. The spectral variation ($E = \hbar\omega = 0.83\ to\ 2.067 eV$) of Faraday rotation $\psi$ (**b,** right axis, red solid line) and ellipticity $\xi$ (**c,** right axis, red solid line) for grating periodicity $d$ =500 nm within the window of the avoided crossing. The spectral variation of the transmitted TM polarized intensity are also shown (left axis, blue dashed line). The simulated spatial distribution of the magnetic field ($H$) for input TM polarization ($H_y$) at energies (frequencies) corresponding to the two transmission dips are shown for grating period $d$ =400 nm (below the avoided crossing regime) (**d**), for $d$ =500 nm (at the avoided crossing regime) (**e**), and for $d$ =650nmabove the avoided crossing(**f**). The cross section of the WMPC is shown by white solid lines, with the rectangles representing the position of the Au gratings.

# Supplementary Information

# Unravelling the role of waveguide-plasmon strong coupling, avoided crossing and natural weak value amplification on the enhanced Faraday effect in hybrid magneto-plasmonic systems

Jeeban K Nayak[+], Shyamal Guchhait[+], Ankit K Singh and Nirmalya Ghosh*

*Department of Physical Sciences,*

*Indian Institute of Science Education and Research (IISER) Kolkata.*

*Mohanpur 741246, India*

+these authors contributed equally

**Corresponding author: nghosh@iiserkol.ac.in*

## ***Supplementary note 1. Theoretical model of natural weak value amplification in Fano resonance***

Our natural weak value amplification (WVA) formalism is founded on a simple but intuitive model of optical Fano resonance that uses coherent interference of a narrow resonance mode described by a complex Lorentzian with a frequency-independent continuum (or broad) mode[1,2]. It has been observed earlier that similar simple model can capture the near field mode coupling and the Fano interference effect in various optical systems including the plasmonic crystals[3–8]. The resultant Fano scattered electric field is given by

$$E_s(\omega) \approx \left[\frac{(q-i)}{(\epsilon+i)} + B\right] = \left[\frac{\sqrt{q^2+1}}{\sqrt{\epsilon^2+1}} e^{i\psi(\omega)} + B\right] \qquad \text{(S1)}$$

The total phase difference between two interfering modes $\Psi(\omega)$ is

$$\Psi(\omega) = tan^{-1}\left[\frac{(q+\epsilon)}{1-q\epsilon}\right] \qquad \text{(S2)}$$

The resulting expression for the transmitted intensity of electric field described in Eq. S1 can be obtained as

$$I_s(\omega) = |E_s(\omega)|^2 = B^2 \times \left[\frac{\left(q^{eff}+\epsilon\right)^2}{(\epsilon^2+1)}\right] + \frac{(B-1)^2}{(\epsilon^2+1)} \qquad \text{(S3)}$$

The parameter $B$ is the relative amplitude of the continuum with respect to the field of the narrow resonance and determines the contrast of the spectral domain interference. For the ideal situation of $B = 1$, the perfect destructive interference occurs at the Fano frequency $\omega_F = \left(\omega_0 - \frac{q\gamma}{2}\right)$ corresponding to $(\epsilon = -q,\ \Psi(\omega_F) = \pi)$[1,2].

Here we are proposing a generalized model of Fano interference in the WMPC by considering different resonant frequencies for TE and TM waveguide modes. The resultant polarized field can be expressed as

$$\boldsymbol{E_s}(\omega) \approx= \left[\frac{q-i}{\varepsilon(\omega)^{TE}+i}(\cos\alpha\cos\chi - i\sin\alpha\sin\chi)\hat{\boldsymbol{y}} + \frac{q-i}{\varepsilon(\omega)^{TM}+i}(\sin\alpha\cos\chi + i\cos\alpha\sin\chi)\hat{\boldsymbol{x}} + B\hat{\boldsymbol{y}}\right] \quad \text{(S4)}$$

Here, $\varepsilon(\omega)^{TE} = \frac{\omega-\omega_0{}^{TE}}{(\gamma/2)}$, $\varepsilon(\omega)^{TM} = \frac{\omega-\omega_0{}^{TM}}{(\gamma/2)}$ , where $\omega_0{}^{TE}$ and $\omega_0{}^{TM}$ are the resonant frequencies of TE and TM waveguide modes respectively and $\gamma$ is the width of the narrow resonance. The q parameter is related to the coupling of modes and determines the characteristic asymmetry of Fano spectral line shape. In this model, the spectral domain interference of the modes can be understood through the phase difference between them, which is very important for our weak value analysis[1,2].

$$\Psi(\omega)^{TE} = tan^{-1}\left[\frac{(q+\epsilon^{TE})}{1-q\epsilon^{TE}}\right]$$

$$\Psi(\omega)^{TM} = tan^{-1}\left[\frac{(q+\epsilon^{TM})}{1-q\epsilon^{TM}}\right] \quad \text{(S5)}$$

$\Psi(\omega)^{TE/TM}$ is the total phase difference between the waveguide mode (TE/TM) and the broad continuum. For an ideal continuum $(B = 1)$, the exact destructive interference occurs at the Fano frequency $\omega_F = \left(\omega_0{}^{TE} - \frac{q\gamma}{2}\right)$ corresponding to $\varepsilon^{TE} = -q$. Note that $\omega_F$ is the frequency of exact destructive interference, where the phase difference between the TE quasiguided mode and incoming broad continuum becomes $\Psi(\omega_F)^{TE} = \pi$ and the ratio of the amplitudes between them is unity $a(\omega_F) = 1$. At close vicinity of $\omega_F$, near destructive spectral interference takes place between the two modes with simultaneous small amplitude offset $\left(\epsilon_a(\omega)\right)$and phase offset $\left(\epsilon_p(\omega)\right)$. It is this near destructive spectral interference between the two modes that mimics the near orthogonal pre and post-selections of states (small overlap of states $\epsilon_{a/p}$) in conventional optical WVA. It's important to note that, in the general case, the amplitude and phase offset parameters for TE $(\epsilon_{a/p})$ and TM $(\epsilon'_{a/p})$ interfering modes will differ owing to their different resonance frequencies, which can be expressed as (expression is given in Eq. 2 and Eq. 3 of the manuscript).

Now the WVA of polarization rotation $\alpha$ and ellipticity $\chi$ will depend on both the offsets $\left(\epsilon_{a/p}{}^{TM},\ \epsilon_{a/p}{}^{TE}\right)$. Using the framework of WVA considering two different amplitude and phase offsets, the resultant Fano scattered electric field described in Eq. 1 can be written in terms of simultaneously varying offset parameters( $\epsilon_{a/p}, \epsilon'_{a/p}$ ). In this complex WVA scenario it can be written as

$$\boldsymbol{E_s}(\omega) = \left[(1+\epsilon_a)e^{+i\epsilon_p}\left\{(\cos\alpha\cos\chi - i\sin\alpha\sin\chi)\hat{\boldsymbol{y}} + (1+\epsilon'_a)e^{+i\epsilon'_p}(\sin\alpha\cos\chi + i\cos\alpha\sin\chi)\hat{\boldsymbol{x}}\right\} - (1-\epsilon_a)e^{-i\epsilon_p}\hat{\boldsymbol{y}}\right] \quad \text{(S6)}$$

Here $\epsilon_a$ and $\epsilon_p$ are the amplitude and phase offsets of the TE waveguide mode and the $\epsilon'_a$ and $\epsilon'_p$ corresponds to the offsets of the TM waveguide mode which can be calculated using Eq. 3 of the manuscript. The simultaneous real and imaginary WVAs of polarization rotation $\alpha$ and ellipticity $\chi$ are manifested in the $\epsilon_{a/p}$ and $\epsilon'_{a/p}$ dependent large observed polarization vector orientation angle $\psi$ (determined from the $U$ and $Q$ Stokes parameters[10]) and the circular (elliptical) polarization descriptor 4th Stokes vector element $\left(\frac{V}{I}\right)$. As the TE and TM waveguide modes moves away from each other one cannot approach a very small value of amplitude and phase offsets for both the waveguide modes $\left(\epsilon_{a/p}{}^{TM} \to 0,\ \epsilon_{a/p}{}^{TE} \to 0\right)$ simultaneously, which is very much desirable for optimal weak value amplification. Consequently, the maximum achievable amplification of polarization rotation $\alpha$ and ellipticity $\chi$ gets limited. In contrast, for the ideal WVA scenario ($\omega_0^{TE} = \omega_0^{TM} = \omega_0$) the conditions of getting very small offsets are met for both the waveguide modes $\left(\epsilon_{a/p}{}^{TM} \to 0,\ \epsilon_{a/p}{}^{TE} \to 0\right)$ around the Fano dip $\omega_F$ leading to the maximum achievable amplification. Some results are presented below using Eq. S4 and Eq. S6 demonstrating the behaviour of natural WVA with varying overlap between the TE and TM waveguide modes.

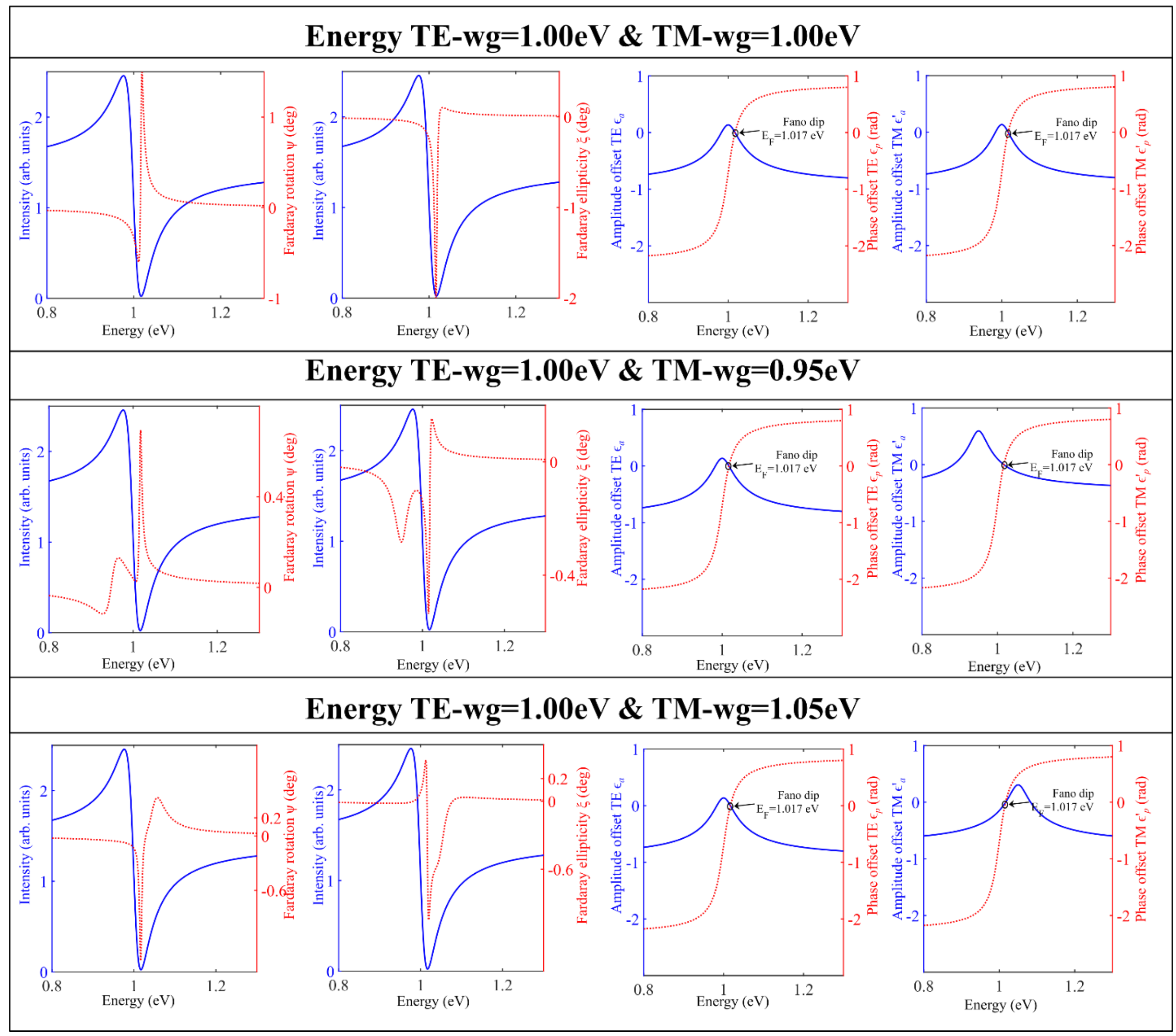

***Supplementary figure 1: Natural weak value amplification in Fano resonance with varying spectral overlap between TE and TM waveguide modes: -*** The theoretical results are shown for demonstrating the behaviour of WVA with varying the spectral overlap between the TE and TM quasiguided waveguide modes. Three different scenarios are chosen with one corresponds to the ideal scenario of WVA ($\omega_0^{TE} = \omega_0^{TM}$) and in others we have different resonant frequencies of TE and TM waveguide modes ($\omega_0^{TE} \neq \omega_0^{TM}$). The spectral variations($E = \hbar\omega = 0.83\ to\ 2.067eV$)of optical rotation $\psi$(right axis, red dashed line) (i) and ellipticity $\xi$ (right axis, red dashed line) (ii) are shown along with the transmitted intensity profiles (left axis blue solid line). ((iii) and (iv)) Spectral variations of the weak measurement parameters, amplitude offset $\epsilon_a$(left axis, blue solid line) and the phase offset $\epsilon_p$(right axis, red dashed line) around the Fano spectral dip ($E = \hbar\omega_F$) for both the waveguide modes. The following parameters of Fano resonance were used for these theoretical calculations: $q = -1$, $B = 1.2$, and $\alpha = 0.28$ deg., $\phi = 0.1deg.$.

It is evident from the results that for scenarios where the TE and TM waveguide modes have different resonance frequencies ($\omega_0^{TE} \neq \omega_0^{TM}$), both the offset parameters $\left(\epsilon_{a/p}, \epsilon'_{a/p}\right)$ are not simultaneously approaching zero around the Fano dip $\omega_F$, which is very much desirable for achieving optimal amplification. Hence the maximum achievable amplification of both the Polarization rotation and ellipticity get limited with increasing the difference between $\omega_0^{TE}$ and $\omega_0^{TM}$ as one can not approach to small offset for both the waveguide modes simultaneously. It is also important to note that the WVA approximation does not break down even if the resonance modes are spectral away from each other, only the amplification value goes down with possibly slight deviation from the ideal WVA behaviour which is known to be more accurate for small magnitudes of the amplitude and the phase offset parameters.

***Supplementary note 2. Magneto-optical figure of merit of the waveguided magneto-plasmonic crystal***

Simultaneous enhancement of both the Faraday rotation and optical transmission is much needed for photonic device applications.  For input TE polarization the enhancement of the Faraday effects is happening around the Fano spectra dip as a direct manifestation of natural WVA and as per all conventional weak measurements it comes with a cost of the total signal. Yet as we have discussed in the manuscript, optimal amplification of Faraday effects and moderate transmittance can be obtained simultaneously by carefully choosing a frequency window around the Fano spectral dip. A figure describing the rotation and transmittance for TE input polarization is given below in which the typical spectral window that can be utilized for this purpose is marked. It may also be noted that here one is dealing with the elastically scattered light from the WMPC system where the low intensity signal is not expected to be a significant issue for all practical purposes. It is also pertinent to note in this regard that in all the actual optical Fano resonant systems (like the WMPC), the scattered intensity never reaches zero at the Fano spectral dip due to the presence of finite dissipation in practical systems. This also enables one to work even at the exact Fano spectral dip, enabling one to utilize the maximum amplification effect.

We have also provided the comparison of magneto-optical figure of merit (MOFOM) between the WMPC and the bare magneto-optical film in the manuscript (Fig. 3(A)), which clearly shows the advantage of using the WMPC and also provides strong evidence of its eligibility for being a potential candidate for building nonreciprocal photonic devices.

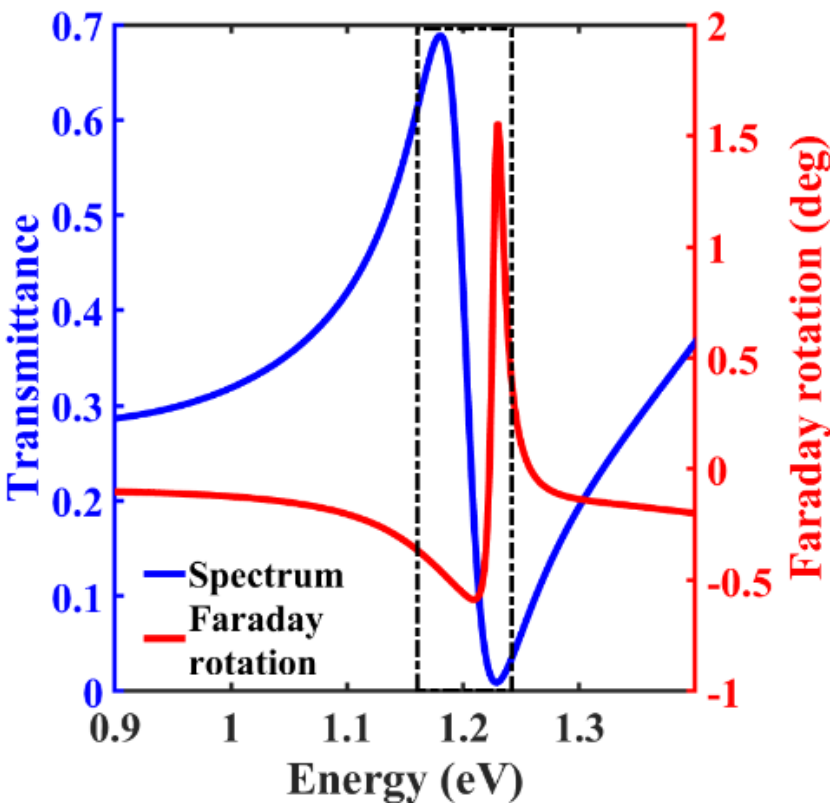

***Supplementary figure 2:*** The Spectral variation of Faraday rotation (red solid line) is plotted along with the transmission spectra (blue solid line). A frequency window around the Fano dip is marked (black dotted line), where we can have both enhanced Faraday effects and moderate transmittance.

***Supplementary note 3. Simulated field distribution for input TE polarization***

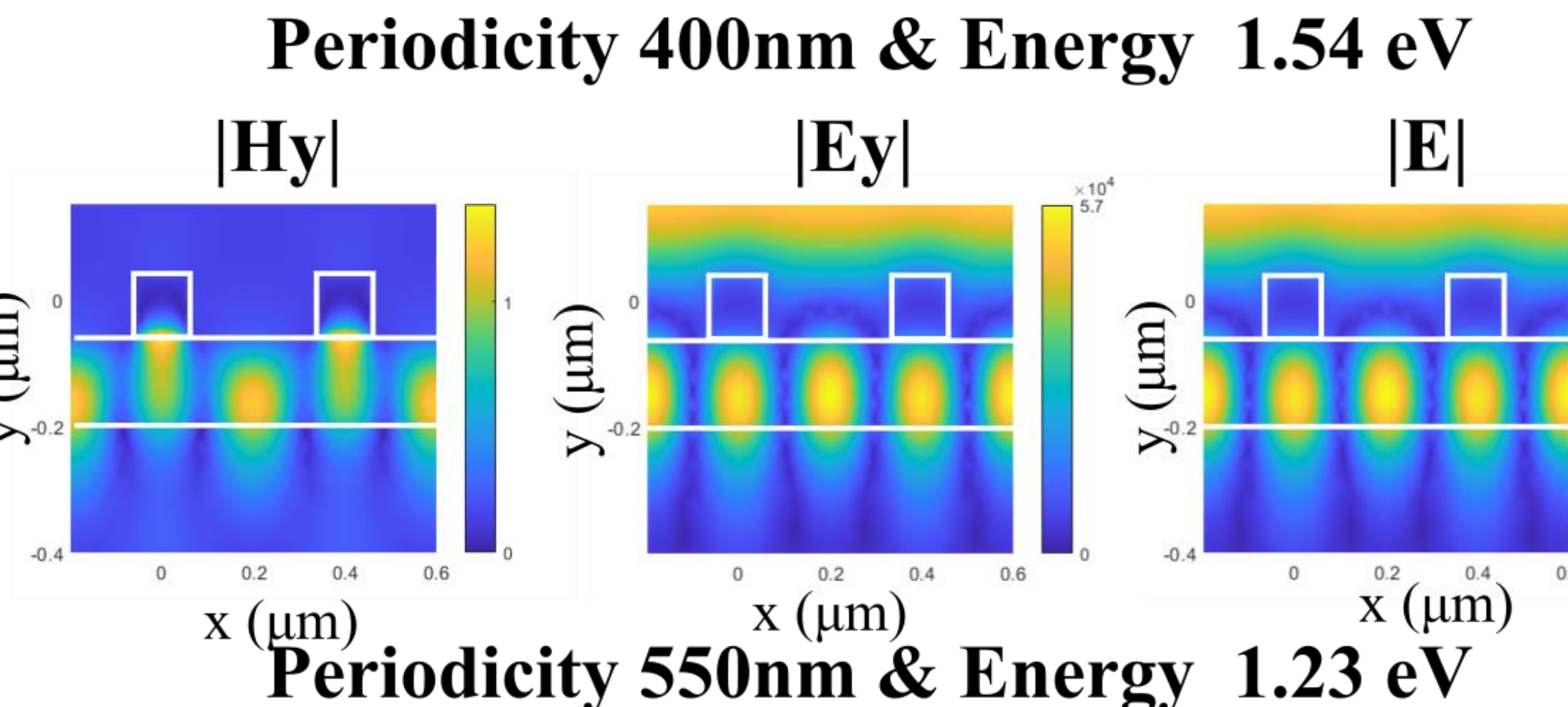

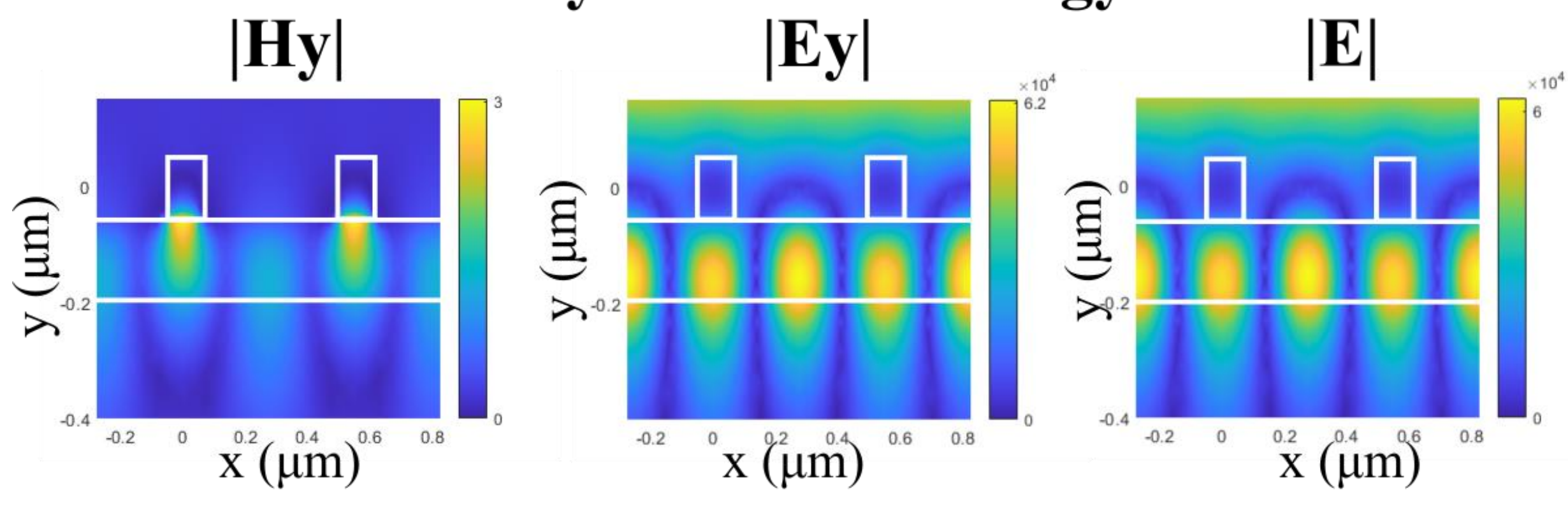

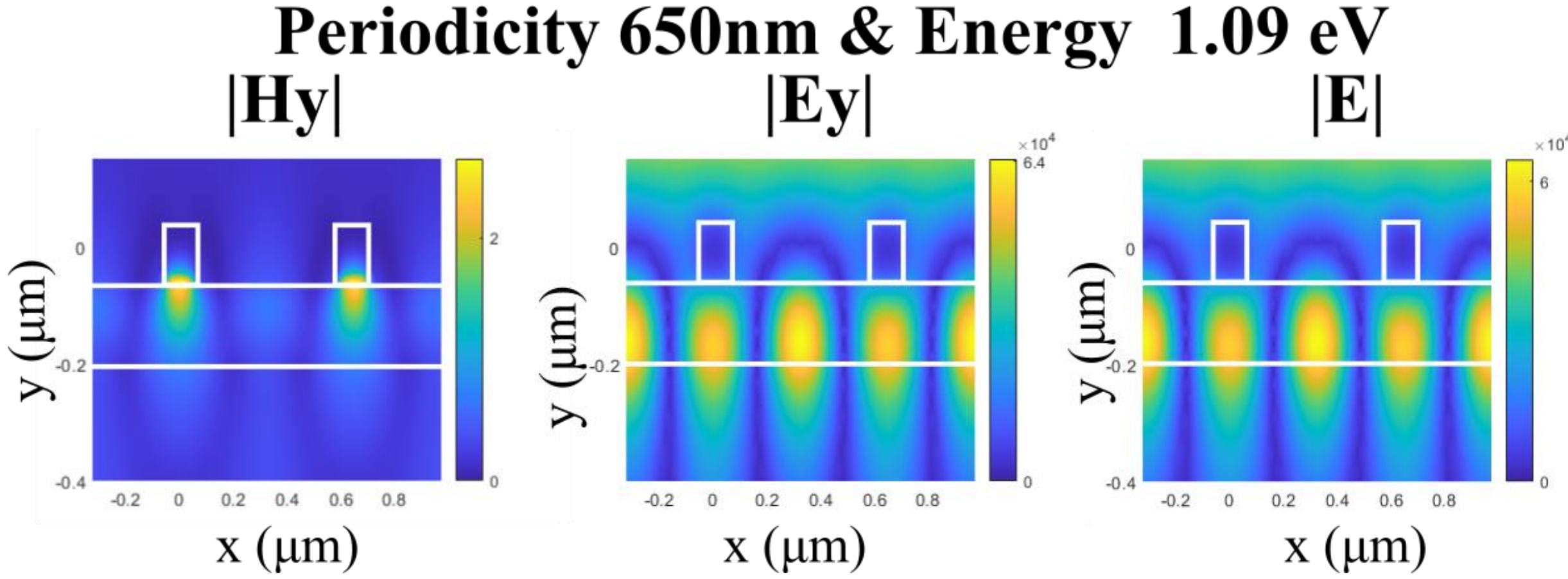

***Supplementary figure 3: Near field distribution around the Fano dip with input TE polarization: -*** The near field spatial distribution of both TE ($|E_Y|$) and TM($|H_Y|$) polarized field are given along with the total field ($|E|$) around the Fano dip for three different periodicities with input TE excitation. The cross section of the WMPC is shown by white solid lines, with the rectangles representing the position of the Au gratings.

From the above figure it is evident that for all the grating periodicities, strength of the TE polarized field is much stronger than TM and both the TE and TM polarized fields are primarily waveguiding in nature. It ensures that for input TE excitation the near destructive spectral domain interference of the Fano resonance leads to the natural WVA of the Faraday effects and the observed enhanced Faraday rotation and ellipticity are direct manifestation of this natural WVA.

**Supplementary References: -**